\documentclass[prb,superscriptaddress,floatfix,twocolumn,showpacs,amsfonts,amsmath,amssymb]{revtex4}
\usepackage{amssymb}
\usepackage{bm}
\usepackage{amsmath}
\usepackage[dvips]{graphicx} 
\usepackage{bmpsize}

\begin{document}

\title{Manifestation of a non-abelian gauge field in a p-type semiconductor system}
\author{T. Li}
\affiliation{School of Physics, University of New South Wales, Sydney 2052, Australia}
\author{L. A. Yeoh}
\affiliation{School of Physics, University of New South Wales, Sydney 2052, Australia}
\author{A. Srinivasan}
\affiliation{School of Physics, University of New South Wales, Sydney 2052, Australia}
\author{O. Klochan}
\affiliation{School of Physics, University of New South Wales, Sydney 2052, Australia}
\author{ D. A. Ritchie}
\affiliation{Cavendish Laboratory, J. J. Thomson Avenue, Cambridge CB3 0HE, United Kingdom}
\author{ M. Y. Simmons}
\affiliation{Centre for Quantum Computation and Communication Technology, School of Physics, University of New South Wales, Sydney NSW 2052, Australia}
\author{ O. P. Sushkov}
\affiliation{School of Physics, University of New South Wales, Sydney 2052, Australia}
\author{ A. R. Hamilton}
\affiliation{School of Physics, University of New South Wales, Sydney 2052, Australia}

\begin{abstract}
Gauge theories, while describing fundamental interactions in nature, also 
emerge in a wide variety of physical systems. Abelian gauge fields have 
been predicted and observed in a number of novel quantum many-body systems, 
topological insulators, ultracold atoms and many others. However, the 
non-abelian gauge field, while playing the most fundamental role in 
particle physics, up to now has remained a purely theoretical construction 
in many-body physics. In the present paper we report the first observation 
of a non-abelian gauge field in a spin-orbit coupled quantum system. 
The gauge field manifests itself in quantum magnetic oscillations of a hole 
doped two-dimensional (2D) GaAs heterostructure. Transport measurements were 
performed in tilted magnetic fields, where the effect of the emergent 
non-abelian gauge field was controlled by the components of the magnetic 
field in the 2D plane.
\end{abstract}
\maketitle

Gauge theories were originally conceived to describe elementary particles 
and their interactions \cite{YangMills, Glashow}. The concept of the emergent 
gauge field is relevant to a wide class of quantum systems whose initial 
formulation has no apparent relationship to gauge fields. Such emergent 
gauge fields arise naturally in many geometrical contexts and the idea that 
physical systems can be classified according to their geometrical properties 
has become an overarching paradigm of modern physics. One example of an 
abelian gauge theory in this context is the Berry phase \cite{berry}, which 
is associated with the adiabatic evolution of
a nondegenerate quantum state. The emergence of non-abelian gauge fields 
in degenerate quantum systems was first theoretically proposed by Wilczek 
and Zee \cite{WZ} shortly after the work of Berry.

While abelian gauge fields have been observed in systems ranging from optical
fibers \cite{tomita86} and semiconductor rings \cite{Shayegan,Wieck} to Bose 
condensates of ultracold atoms \cite{Lin09}, signs of non-abelian effects 
have so far only been observed in the nuclear quadrupole resonance of 
$^{35}$Cl in a single crystal of sodium chlorate \cite{zwanziger}. Non-abelian 
gauge fields have been theoretically predicted in a number of many-body 
systems including fractional quantum Hall liquids~\cite{moore}, spin-orbit 
coupled systems~\cite{arovas,Murakami}, cuprate superconductors~\cite{Lee} 
and ensembles of ultracold atoms~\cite{Dalibard,gerbier}.
In spite of the theoretical excitement and great interest
all previous attempts to observe these fields were unsuccessful. This 
demonstrates the challenge involved in the experimental realization of 
emergent non-abelian gauge fields.

The idea of our experiment is partially based on previous theoretical work 
by Arovas and Lyanda-Geller \cite{arovas} as well as Murakami, Nagaosa 
and Zhang \cite{Murakami} who proposed that effects relating to non-abelian 
gauge fields must be pronounced in hole-doped zinc blende semiconductors due 
to the strong spin-orbit coupling (SOC). In this context the gauge fields 
are closely associated with spin dynamics along curved trajectories: 
Ref.~\cite{arovas} proposed the use of mesoscopic rings to bend the 
trajectory, whilst Ref.~\cite{Murakami}
suggested use of an external electric field for the same purpose. In this 
work we use a 2D GaAs hole-doped heterostructure in a relatively small 
(fraction of a Tesla) magnetic field applied perpendicular to the 2D plane 
to curve the hole trajectories. In addition, we apply an in-plane magnetic 
field
($B_{||}\sim $ several Tesla), which allows us to control the magnitude of the
spin-orbit coupling.
The combination of the SOC and curved trajectories
makes the non-abelian gauge field  observable.
The perpendicular magnetic field gives rise to quantum magnetic oscillations
which are influenced by non-abelian spin dynamics. We measure the oscillations
via the Shubnikov-de Haas (SdH) effect.
{
The SdH effect has been measured previously in numerous experiments with
2D systems with strong spin-orbit interaction, see e.g. Refs.~\cite{Engels,Grbic}
However in all previous studies, effects related to the non-abelian Berry phase are 
negligible, and what is measured is simply the densities of the spin-split subbands.
One needs very special conditions to distinguish between the  abelian and
the non-abelian Berry phases, it is necessary to tune independently
the spin precession, the orbital dynamics, and the spin-orbit interaction.
To do so in our experiment we use the following crucial points.
}
(i) We can tune the spin-orbit coupling over a wide range using the in-plane 
field $B_{||}$ while keeping the orbital dynamics fixed.
(ii) We use a low symmetry crystal with highly anisotropic coupling to $B_{||}$,
which allows us to control independently the Larmor and the spin precession
frequencies. This is key 
to proving that the effects we observe cannot be due to abelian physics, nor 
due to differences between datasets taken at different carrier densities, 
gate biases, or even from different samples.
(iii) We use a device where we can minimise the undesirable Rashba
interaction, allowing a simple analytic theory to explain the data.
These three factors allow us to report the first observation
of the non-abelian gauge phase which was elusive for
30 years since its theoretical prediction.

\begin{figure*}[ht!]
\begin{tabular} { c c  c}
\Large {
{\bf a}  } & \Large {{\bf b} } & \Large {{\bf c} } \\
\includegraphics[width = 0.25\textwidth]{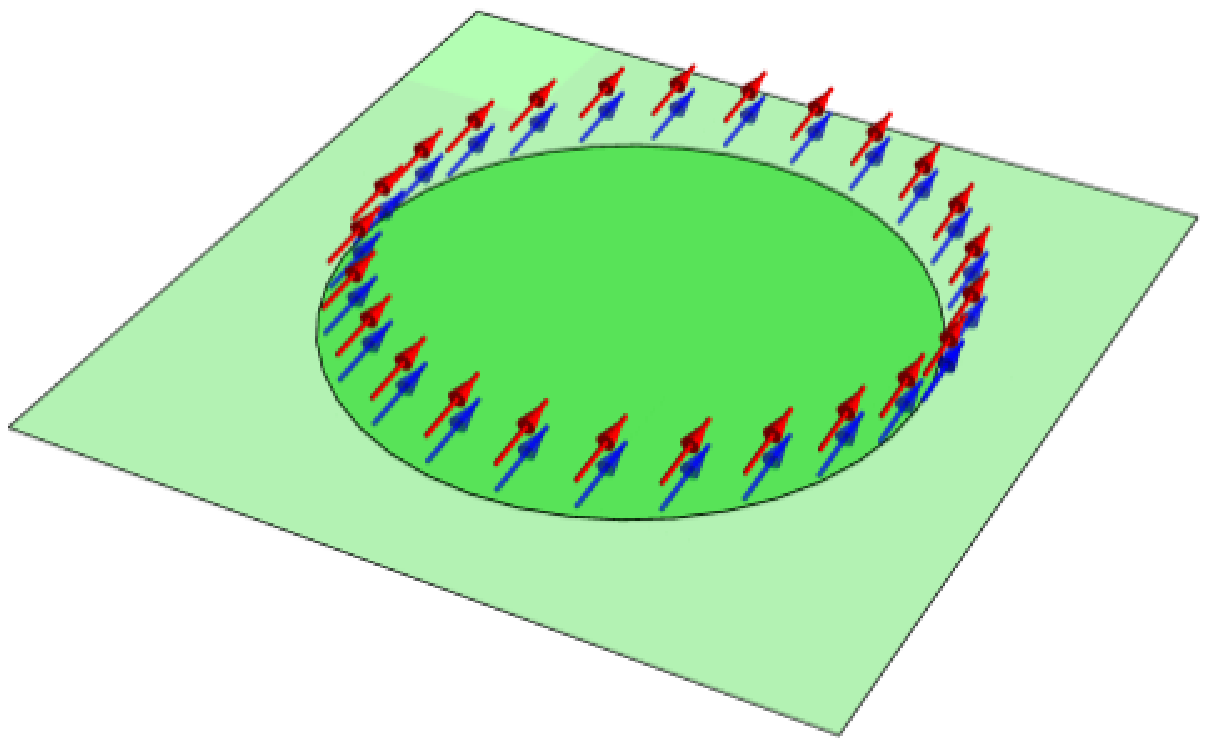} &
\includegraphics[width = 0.25\textwidth]{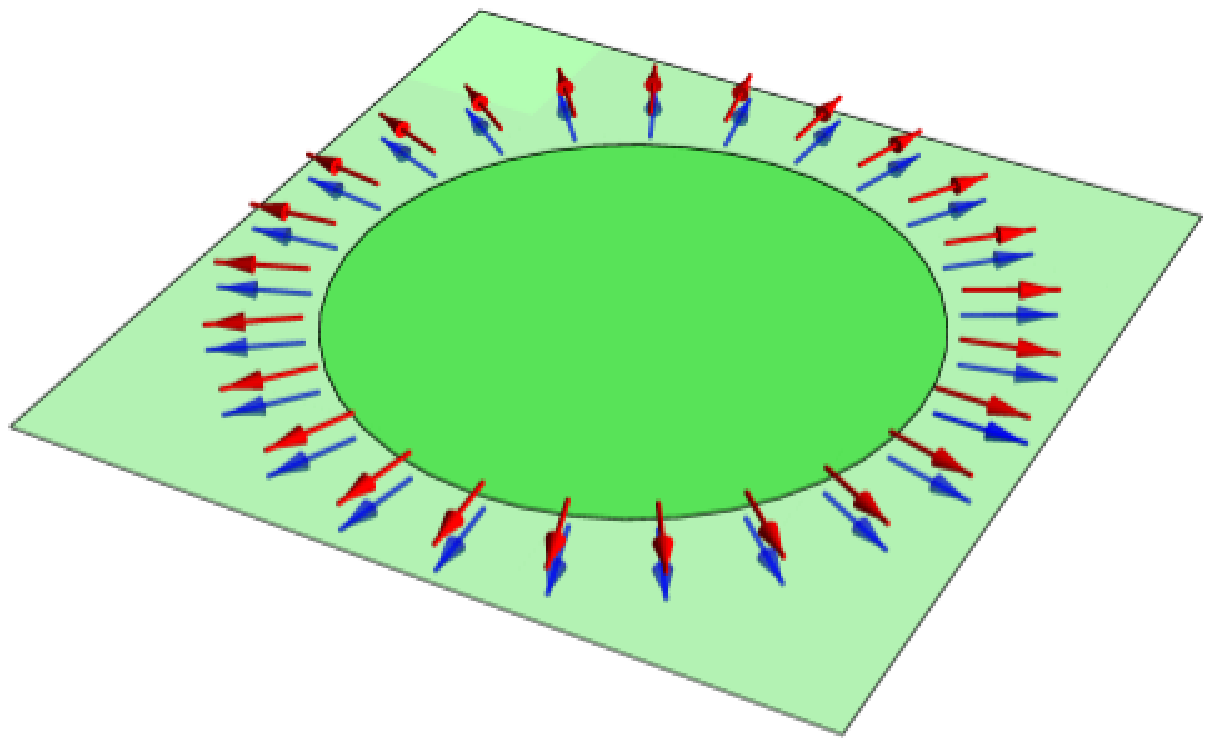}&
\includegraphics[width = 0.25\textwidth]{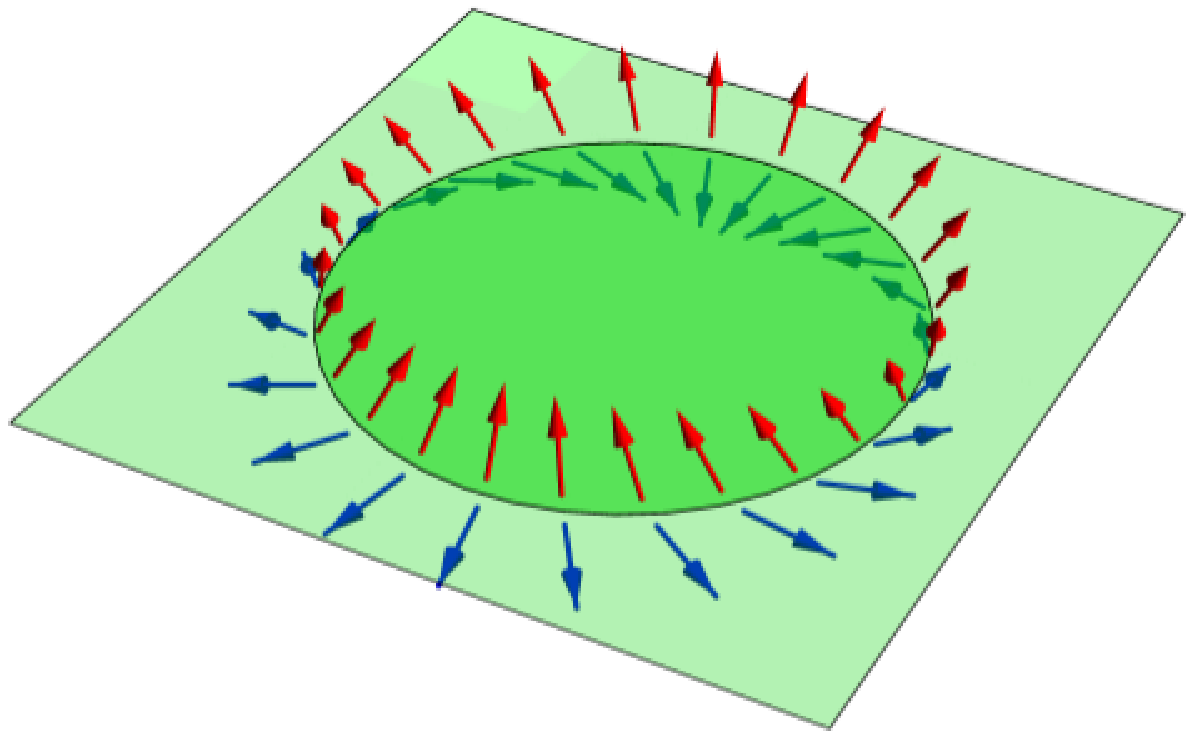}
\end{tabular}
\caption{Spin dynamics along the closed trajectory in momentum space (green 
circle) in three qualitatively different situations. The spin shown by red 
arrows is driven by a local effective magnetic field shown by blue arrows. 
{\bf (a)} Absence of spin dynamics. This corresponds to the case of an 
electron moving within an external magnetic field in the absence of 
spin-orbit interaction. {\bf (b)} Abelian spin dynamics. Spin is changing, 
but it remains parallel to the driving field 
$\boldsymbol{B}_{eff} \propto \boldsymbol{k}$. {\bf (c)} Non-abelian spin 
dynamics. The spin is parallel to the vector sum of the driving field 
$\boldsymbol{B}_{eff}$ and the non-abelian gauge field.}
\label{fig1}
\end{figure*}

The spin dynamics of a particle moving around a circle in momentum space
used in Onsager quantization \cite{Onsager}
is illustrated in Fig.\ref{fig1}. The three panels in this figure correspond 
to three qualitatively different situations: (a) spin dynamics being absent, 
(b) abelian spin dynamics and (c) non-abelian spin dynamics. The spin 
(red arrows) is driven by a local effective magnetic field 
$\boldsymbol{B}_{eff}$ (blue arrows), which is the sum of the external 
magnetic field $\boldsymbol{B}_{ext}$ and the momentum-dependent spin-orbit 
field
$\boldsymbol{B}_{soc}$. Panel (a) depicts the trajectory of a non-relativistic 
electron in the absence of spin-orbit. In this case 
$\boldsymbol{B}_{eff} = \boldsymbol{B}_{ext}$ and the spin is simply aligned  
with the external field. Panel (b) illustrates the case of an ultrarelativistic
 Dirac electron, \emph{e.g.} an electron in graphene or on the surface of a 
topological insulator. In this case, although  spin is precessing, it remains 
aligned with the driving field which itself is parallel to the momentum, 
$\boldsymbol{B}_{eff}\approx \boldsymbol{B}_{soc} \propto {\bm k}$. The 
precession of spin around the orbit generates a geometric Berry phase 
(abelian gauge field)
which appears as the $\pi$-phase shift observed in magnetic oscillations 
\cite{graphene,ti}.
The non-abelian case addressed in the present work is illustrated in panel (c).
 Here, the driving field $\boldsymbol{B}_{eff}$ is not collinear with spin and 
the noncollinearity is proportional to the non-abelian gauge field. Due to 
the non-abelian spin dynamics, the particle acquires a matrix-valued phase 
equal to the circulation of the gauge field around the trajectory in momentum 
space. The phase manifests itself in quantum magnetic oscillations.

Holes in GaAs originate from atomic $p_{3/2}$ orbitals and hence posses an 
angular momentum $J=3/2$. The electric quadrupole interaction leads to strong 
coupling between the angular momentum ${\bm J}$ and the linear momentum 
${\bm k}$, which is described by the  Luttinger Hamiltonian \cite{Luttinger}. 
The $z$-confinement in a
2D heterostructure enforces  quantization of ${\bm J}$ along the $z$-axis. 
Therefore, a hole quantum state with a given in-plane momentum 
${\bm k}=(k_x,k_y)$ splits into two doublets with $J_z=\pm 3/2$ (heavy holes) 
and $J_z=\pm 1/2$ (light holes). Since light holes lie significantly higher 
in energy, we shall only consider heavy holes for the low energy dynamics.

The heavy-hole Kramers doublet can be described by an effective spin $s=1/2$,
$|J_z=+3/2\rangle \equiv|\uparrow\rangle$,
$|J_z=-3/2\rangle \equiv |\downarrow\rangle$.
The Hamiltonian describing heavy holes consists of
the kinetic energy, the Zeeman interaction and the SOC,
$H=H_K+H_Z+H_{soc}$,
\begin{eqnarray}
\label{Heff}
&&H_K=\epsilon({\bm k}) \\
&&H_Z=-\frac{\Delta}{2}\sigma_z \ , \ \ \Delta=g\mu_B B_z \nonumber\\
&&H_{soc}\equiv
- \boldsymbol{\beta}(\boldsymbol{k}) \cdot \boldsymbol{\sigma}=
-\frac{1}{2}\alpha\left[\sigma_+B_-k_-^2+\sigma_-B_+k_+^2\right]   ,\nonumber
\end{eqnarray}
where ${\bm k}=-i\hbar{\bm \nabla}-e{\bm A}$; 
$\sigma_{\pm}=\sigma_x\pm i\sigma_y$, $B_{\pm}=B_x\pm iB_y$, $k_{\pm}=k_x\pm ik_y$; 
${\bm A}$ is the in-plane vector potential 
created by $B_z$, $e$ is the elementary charge, $\sigma_i$ are Pauli matrices 
describing the spin, $\mu_B$ is Bohr magneton, 
$g=g_{zz}$ is the effective g-factor and $\alpha$ is 
the SOC strength.
Note that due to mixing between heavy holes states the dispersion
$\epsilon(k)$ can significantly differ from the simple quadratic
form, see discussion in  Appendix \ref{apD}.
Note also that generally $g$ and $\alpha$ depend on $k$, and
in combination with nonquadratic dispersion $\epsilon(k)$ this dependence 
results in a very complex fan diagram of Landau levels.
However, according to the Landau theory of normal Fermi liquids this complexity is 
irrelevant to
the problem we address. We do not need the full Landau level fan diagram.
According to normal Fermi liquid theory only the values of the parameters at the
Fermi energy are relevant. This statement is very general, and even includes hole-hole
Coulomb interaction effects. We will fit the experimental data to obtain the
parameters $g$  and $\alpha$ at $\epsilon=\epsilon_F$.
A derivation of
the spin-orbit interaction $H_{soc}$ is presented in 
Appendix \ref{apD}, although we shall make two comments here on its origin: 
(i) The spin-orbit coupling arises from a small mixing between heavy and 
light holes, where the mixing probability is 1-2\% (see Appendix \ref{apD}). 
(ii) The kinematic structure of $H_{soc}$ in Eq.(\ref{Heff}) is dictated by 
the fact that the Pauli matrices $\sigma_{\pm}$ correspond to $\Delta J_z=\pm 3$.

If the perpendicular magnetic field is zero, ${\bm A}\propto B_z=0$,
then the hole trajectories are straight lines and $H_{soc}$ in Eq.(\ref{Heff}) 
simply splits the doubly degenerate band, $\epsilon_k$, into a pair of 
chiral bands.
In presence of $B_z$
the hole trajectory forms a circle,
${\bm k}=k(\cos\theta,\sin\theta)$.
Semiclassically, using the wave packet picture, the angle is
$\theta=-\omega_c t$ (the sign corresponds to $B_z > 0$), where
$\omega_c=e|B_z|/m$ is the cyclotron frequency
and $m=k\left(\frac{d\epsilon}{dk}\right)^{-1}$ is the effective 
cyclotron mass at the Fermi energy.
The spin-orbit field 
$\boldsymbol{\beta}(\boldsymbol{k})$ varies along the trajectory. This 
variation can be removed by a local gauge transformation of the spinor wave 
function $\psi\to \psi^{\prime}=g^{-1}(\boldsymbol{k})\psi$. Taking 
$g(\boldsymbol{k}) = e^{ - i \theta\sigma_z }$ we gauge out the angle 
dependence of the SOC,
\begin{equation}
\label{b1}
\boldsymbol{\beta}^{\prime}(\boldsymbol{k})\cdot \boldsymbol{\sigma} =
g^{-1} \left[\boldsymbol{\beta}\cdot \boldsymbol{\sigma}\right]g\ , \ \ \
\boldsymbol{\beta}^{\prime} = \alpha k^2  \boldsymbol{B}_\parallel \ .
\end{equation}
Since our choice of  $g(\boldsymbol{k}) $ ensures that we perform a 
transformation to the co-rotating frame of the hole, it follows that 
$\boldsymbol{\beta}^{\prime}$ does not vary along the trajectory. The gauge 
transformation results in the covariant derivative
$\boldsymbol{\nabla} \to \boldsymbol{\nabla} - i \boldsymbol{\Omega}_{\boldsymbol{k}}$, where
$\boldsymbol{\Omega}_{\boldsymbol{k}}$ is the non-abelian gauge field possessing 
a vortex structure in 2D momentum space
\begin{equation}
\label{NAGF}
\boldsymbol{\Omega}_{\boldsymbol{k}} = i g^{-1} \boldsymbol{\nabla}_{\boldsymbol{k}} g=
\left(- \frac{ k_y \sigma_z}{k^2}, \frac{ k_x \sigma_z}{k^2} \right)  \ .
\end{equation}
The field tensor corresponding to this gauge field is zero,
$F_{\mu\nu}=\partial_{\mu}\Omega_{\nu}-\partial_{\nu}\Omega_{\mu}-i[\Omega_{\mu},\Omega_{\nu}]=0$.
However, the gauge field has a nonzero circulation along the hole trajectory
\begin{equation}
\label{flux}
 \oint{ \boldsymbol{\Omega}_{\boldsymbol{k}} \cdot \boldsymbol{dk} } = 2\pi \sigma_z \ \ ,
\end{equation}
and this circulation reveals itself in quantum magnetic oscillations.

To understand quantum magnetic oscillations we need to consider the impact of 
spin-orbit coupling upon the Landau level structure. For this analysis, we 
restrict ourselves to a semiclassical approximation, where the Landau levels 
are determined by the Onsager quantization condition. Consider a hole 
traversing the circular trajectory, where the hole is initially prepared in a 
polarization state $\psi(0)$. Under the combined action of $H_{soc}$ and $H_Z$
spin will precess along the trajectory, as shown in Fig.\ref{fig1}c.
After a full cycle the spin wave function is
$\psi(2\pi) = U \psi(0)$, where $U \in \text{SU}(2)$ is a unitary evolution 
matrix.
In order to
satisfy the semiclassical quantization condition, it is necessary for $\psi(0)$ to be an eigenvector of $U$, i.e.
$\psi(2\pi)=e^{\pm i \Phi}\psi(0)$. Here $e^{\pm i \Phi}$ are the complex conjugate 
eigenvalues of $U$. Hence, depending on the spin state, an additional phase 
$\pm \Phi$ appears in the Onsager quantization condition due to spin dynamics.

SdH oscillations in the resistivity are given by the usual Lifshitz-Kosevich 
formula \cite{LifshitzKosevich}. Accounting for the additional phase $\Phi$ 
we obtain,
\begin{eqnarray}
\Delta \rho_{xx} =
\rho_{xx}(B) - \rho_{xx}(0) =
\mathcal{A}(B) \cos \Phi  \cos 
\frac{ \pi k_F^2}{e|B_z|}
  \ \ .
 \label{osc}
\end{eqnarray}
The amplitude factor depends on the hole 
scattering time $\tau$,
$\mathcal{A}(B) \propto e^{ - \frac{ \pi}{\omega_c \tau}}$.
Spin dynamics enters only via the spin evolution phase factor 
$ \text{tr} U =2 \cos \Phi$. For the semiclassical approximation approach 
we assume large filling factors 
$\nu = \frac{ k_F^2}{2e|B_z|} \gg 1$,
hence 
only the lowest harmonic of magnetic oscillations is taken into consideration.

The matrix phase $U$ may be explicitly expressed as a path-ordered exponential
which can be calculated using the gauge transformation from Eq.(\ref{b1})
\begin{eqnarray}
\label{po}
U&& = \mathcal{P} \exp\left\{ -\frac{i}{\omega_c} \oint\left[
{ \boldsymbol{\beta}} \cdot \boldsymbol{\sigma} + \frac{\Delta}{2} \sigma_z \right] d\theta \right\}
\nonumber\\
&&=
\exp\left\{i\oint{ \Omega_{\boldsymbol{k}} \cdot \boldsymbol{dk} } - i\frac{ 2\pi}{\omega_c} \left[
\boldsymbol{\beta}^{\prime}  \cdot \boldsymbol{\sigma} + \frac{ \Delta}{2} \sigma_z \right] \right\} \  .
\end{eqnarray}
Hence, using Eqs. (\ref{flux}) and (\ref{b1})
we find the prefactor in Eq.(\ref{osc}) for SdH oscillations, $2 \cos \Phi=\text{tr} U$,
\begin{eqnarray}
\Phi
 = \frac{ 2\pi}{\omega_c} \sqrt{ \left(\omega_c - \frac{ \Delta}{2}\right)^2
+ |\alpha k_F^2|^2 (B_x^2 + B_y^2) } \  .
 \label{prefactor}
\end{eqnarray}
Here the $\omega_c$ term under the square root 
comes from the non-abelian gauge field.
It is worth noting that the effect of the gauge field is somewhat
analogous to
Thomas precession in special relativity \cite{Thomas}. As previously 
mentioned, the gauge field cannot be observed without the in-plane magnetic 
field. This is evident from Eq.(\ref{prefactor}): if $\boldsymbol{B}_\parallel=0$
the gauge contribution is exactly $2\pi$ and hence
the phase shift is determined only by the Zeeman splitting, 
$\text{tr} U =2\cos(\pi\Delta/\omega_c)$.
The Zeeman splitting with $\boldsymbol{B}_\parallel\ne 0$ is
$\delta E_Z = \sqrt{ (\frac{ \Delta}{2})^2 + |\alpha k_F^2|^2 (B_x^2 + B_y^2) }$.
A na\"{i}ve expectation for the spin accumulated phase would be
$\Phi_{naive}=2\pi\delta E_Z/\omega_c$, but Eq.(\ref{prefactor}) is different 
from this.
A semi-na\"{i}ve expectation would take into account the abelian Berry phase 
$\varphi_B$ on top of the Zeeman splitting. The phase $\varphi_B$ is given 
by the first term of the square root expansion in (\ref{prefactor})
in powers of $\omega_c$, yielding
\begin{equation}
\label{preB}
\Phi_{AB}=\frac{2\pi\delta E_Z}{\omega_c}+\varphi_B=
\frac{2\pi\delta E_Z}{\omega_c}-\frac{\pi \Delta}{\delta E_Z} \ .
\end{equation}
The subscript ``AB'' in $\Phi$ stands for ``Abelian Berry''.
The abelian Berry phase approach provides a good description for 
magneto-oscillations in Dirac fermion systems \cite{graphene, ti}, and for 
quantum interference in mesoscopic rings with strong spin-orbit 
coupling \cite{Shayegan, Wieck}. However in our case, both the 
``na\"{i}ve'' $\Phi_{naive}$ and the abelian Berry phase $\Phi_{AB}$ approach 
are inconsistent with the data.

\begin{figure*}[ht!]
\includegraphics[width = 0.8\textwidth]{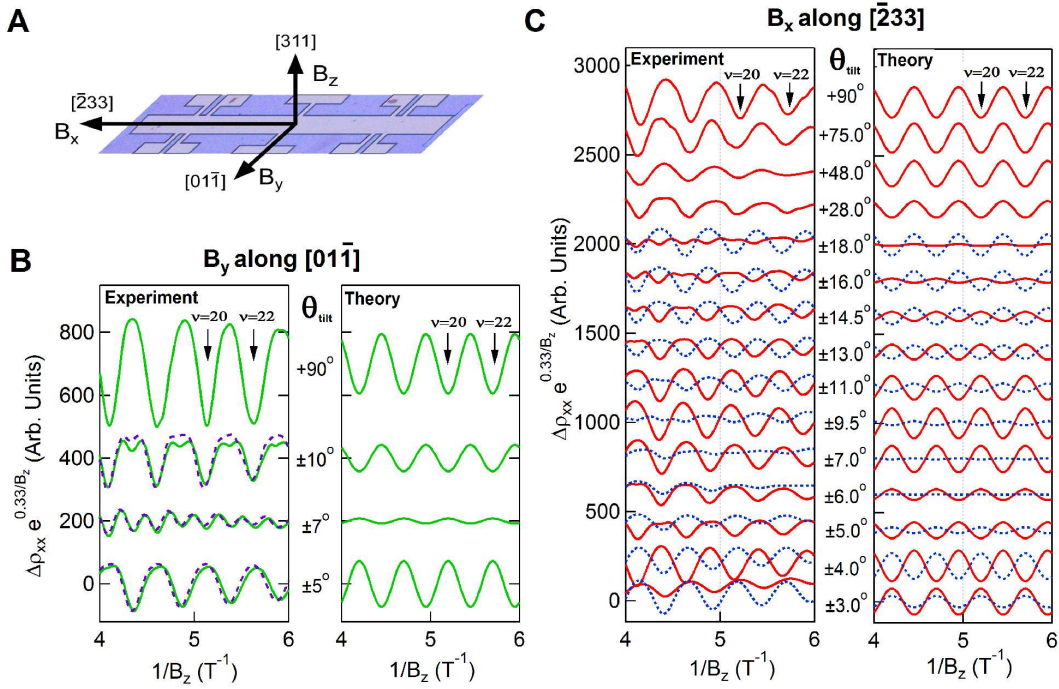}
\caption{
Magnetoresistance (SdH) oscillations in an external magnetic field  
$\boldsymbol{B}$ applied at an angle $\theta_{tilt}$ to the 2D heterostructure, 
$B_{\parallel}=B_z\tan\theta_{tilt}$, $B_z > 0$. {\bf (a)} The orientation of the 
magnetic field components relative to the crystal axes. {\bf (b)} SdH 
oscillations plotted as a function of $\frac{1}{B_z}$ for an applied field of 
$B_y > 0$, $B_x=0$ (green solid lines) and $B_y < 0$, $B_x =0$ (purple dotted 
lines). Traces are offset for clarity. The amplitude of the oscillations was 
normalized by multiplying the data by $e^{0.33/B_z}$. SdH data is presented for 
the range in which the amplitude of the SdH oscillations is not too large 
($\Delta \rho_{xx} < \rho_{xx}$) and $B_z$ is small enough that we do not 
enter the quantum Hall regime,
see Fig.\ref{M1}.
At $\theta_{tilt} = \pm (7. \pm 0.5^{\circ})$ the oscillations exhibit an 
inversion corresponding to the change in sign of $\cos\Phi$ in Eq.(\ref{osc}). 
{\bf (c)} SdH oscillations where the applied field is $B_x > 0$, $B_y=0$ 
(red solid lines) and $B_x < 0$, $B_y=0$ (blue dotted lines). Due to 
crystallographic anisotropy, the oscillations are distinctly different for 
different signs of $B_x$. In this orientation, the oscillations invert at 
angles 
$\theta_{tilt} = 18 \pm 1^{\circ}, 5.5 \pm 0.5^{\circ}, 3.5 \pm 0.25^{\circ}$ for $B_x > 0$ and
$\theta_{tilt}=-6.5 \pm 0.5^{\circ}$ for $B_x < 0$. The filling factors $\nu$ 
are indicated by arrows at the tops of panels b and c.
The right hand panels in {\bf b} and {\bf c} display theoretical SdH curves 
calculated using the non-abelian theory and the usual Lifshitz-Kosevich 
formula, valid in the regime $\Delta \rho_{xx} \ll \rho_{xx}$.
}
\label{fig2}
\end{figure*}

In our experiments the 2D hole system is formed in a 20nm wide symmetric GaAs 
quantum well, grown in a (311)A
GaAs-Al$_{0.33}$Ga$_{0.67}$As heterostructure as indicated in Fig.\ref{fig2}a.
Previous experiments on this system have shown that holes in a (311)  oriented
quantum well have a tensor g-factor with an unusual off-diagonal term 
$g_{xz}$~\cite{Yeohgxz}. Although tilted field measurements revealed the presence of the $g_{xz}$
term, no comparison of the Shubnikov de Haas oscillations with theory was
possible, as there was no theory available for 2D hole systems in tilted 
magnetic fields. We are now able to show that there is excellent qualitative 
agreement between the experimental data and the new theoretical model based
on the non-abelian   gauge field.

We use the coordinates $x\parallel[\bar{2}33], y\parallel[0 \bar{1}1], z\parallel[311]$ shown in 
Fig.\ref{fig2}a.
The gyromagnetic tensor is not diagonal in the $x$,$y$ and $z$ axes, therefore
 the expression for $\Delta$ presented in Eq.(\ref{Heff}) and used elsewhere 
is now replaced by
\begin{equation}
\label{gg}
\Delta = \mu_B(g B_z  + g_{xz} B_x)  \ \  .
\end{equation}
Note that the off-diagonal tensor component $g_{xz}$
makes the magnetic response different for three orientations of $B_{||}$:
$B_{||}=B_x$, $B_{||}=-B_x$, and $B_{||}=B_y$~\cite{Yeohgxz}.
This triples the amount of data we can get from the same sample.
Details of our experimental setup/method are presented in
Appendixes \ref{apA}, \ref{apC}  and \ref{apB}.

So far we have only considered the effect of the external magnetic field, 
however
spin dynamics can also be influenced by additional couplings, such as the 
Rashba
interaction (stemming from the asymmetry of the interface) and the Dresselhaus
interaction (arising from the lack of inversion symmetry in the bulk GaAs crystal).
We apply a voltage bias to the back-gate, to tune the symmetry of the GaAs quantum well such that
the Rashba interaction is practically zero, see Appendix \ref{apB}.
The Dresselhaus interaction is relatively weak, nevertheless it is important
in some regimes. Moreover, as we discuss below, it brings an additional
confirmation of the non-abelian dynamics.

The results of our measurements are presented in panels b and c of 
Fig.\ref{fig2} which plot resistivity versus $1/B_z$ where the in-plane 
field is altered by tilting the sample at an angle $\theta_{tilt}$ with 
respect to the applied field, such that
$B_z = B_{\parallel}\tan\theta_{tilt}$, with $B_z > 0$.
Panel b corresponds to tilting in the $yz$-plane ($B_x=0$) and panel c 
corresponds to tilting in the $xz$-plane ($B_y=0$). The data in panel b is 
symmetric with respect to $B_y \to -B_y$, whilst the data in panel c exhibits 
asymmetry with respect to $B_x \to -B_x$ due to nonvanishing $g_{xz}$ in 
Eq.(\ref{gg}).
According to Eqs. (\ref{osc}) and (\ref{prefactor}), the normalized amplitude 
of resistivity oscillations, $\cos\Phi$, is a function only of $\theta_{tilt}$ 
and is independent of the magnitude of the total magnetic field 
$\boldsymbol{B}$.
At tilt angles corresponding to changes in the sign of 
$\text{tr} U =2\cos\Phi$, the first harmonic of the SdH oscillations 
invert (i.e. maxima become minima, and vice versa). At these ``coincidence'' 
angles the phase $\Phi$ must coincide with a half-integer multiple of $\pi$.
In the data, these coincidences are observed at the tilt angle
$\theta_{tilt} =\pm(7 \pm 0.5^{\circ})$ for the field applied along the
$yz$-plane in Fig.\ref{fig2}b.
For the field applied in the $xz$-plane (Fig.\ref{fig2}c) there are multiple 
coincidence angles at
$\theta_{tilt} = 18 \pm 1^{\circ}, 5.5 \pm 0.5^{\circ}, 3.5 \pm 0.25^{\circ}$ for $B_x >0$ and
only a single coincidence at $\theta_{tilt}=-6.5 \pm 0.5^{\circ}$ for $B_x < 0$.
The coincidence angles  are plotted in Fig.~3, and are described by 
Eqs. (\ref{prefactor}) and (\ref{gg}).
There are three independent device-specific parameters in these equations, 
which are
$g m$, $2\alpha k_F^2/(g\mu_B)$, and $g_{xz}/g$.
We use the value $m=0.25m_e$ derived in Appendix \ref{apD}
as our reference point and hence we are left with unknowns 
$g$, $\lambda=2\alpha k_F^2/\mu_B$, and $g_{xz}$
which we treat  as free fitting parameters.

\begin{figure*}[ht!]
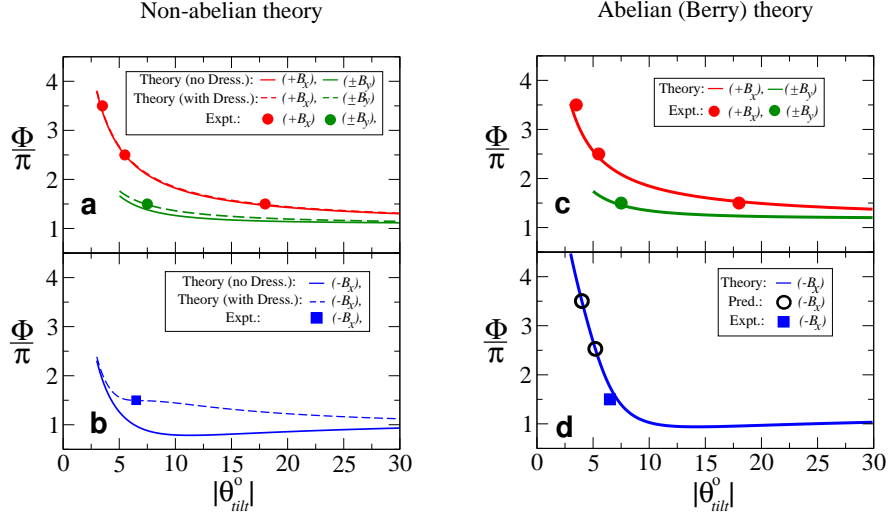

\centering
\begin{tabular}{ c c }
 \includegraphics[width = 0.3\textwidth]{fig3ab.eps}
\hspace{20pt}
 \includegraphics[width = 0.3\textwidth]{fig3cd.eps}
\end{tabular}
\caption{
Comparison of experimentally observed coincidences angles to the non-abelian 
(a, b) and abelian (c, d) theories.
Expressing the envelope of the SdH oscillations as $\cos \Phi$, we find that 
$\Phi/\pi$ becomes a smooth function of tilt angle. This function is plotted 
for the experimental range of tilting angles. Angles at which $\Phi$ crosses 
a half-integer multiple of $\pi$ correspond to inversions of the SdH 
oscillations. Both theories contain three unknown parameters 
$g, g_{xz}, \lambda$. In panels a, c we show the least squares fits of 
$\Phi$ to the observed coincidence angles for an applied $B_x > 0, B_y=0$ 
(red) and $B_y\ne 0, B_x=0$ (green). The plots of $\Phi$ in panels b, d show 
the predicted coincidences for $B_x < 0, B_y=0$. For the non-abelian theory 
the solid curves do not include the Dresselhaus perturbation, whereas it is 
included for the dashed curves. In the abelian case the influence of the 
Dresselhaus perturbation is negligible.
The non-abelian gauge theory predicts a single angle of coincidence  blue 
solid line in panel b, consistent with the observed coincidence point 
(blue square). Including the Dresselhaus interaction (blue dashed line) 
provides quantitative agreement with experiment. In contrast, the abelian 
theory (panel d) predicts three coincidences whilst experimentally only one 
angle of coincidence was observed (solid blue square); the two coincidence 
angles marked  with open symbols were not observed in experiment.
}
\label{fig3}
\end{figure*}

Altogether we have three fitting parameters to describe five coincidence 
angles. To compare the experimental coincidence angles to those of theory, 
we perform a least squares fit to $\Phi$ using the observed four coincidences 
angles for the orientations $B_x > 0, B_y=0$ and $B_y\ne 0, B_x=0$  (red and 
green symbols in Fig. 3a), and use the values obtained to predict the 
coincidence angles for the orientation $B_x < 0, B_y=0$ 
(blue traces in Fig.~3b).
The solid red and green curves in Fig.\ref{fig3}a show the calculated 
$\Phi/\pi$ obtained from this fitting, with the following values of the 
fitting parameters: $g = 7$,
$|\lambda| = 0.88$, and $g_{xz} = -0.87$. The solid blue line in Fig.~3b 
shows $\Phi/\pi$ for the $B_x < 0, B_y=0$ orientation calculated using the 
fitting parameters from Fig.~3a, which predicts that there will be only a 
single coincidence observed in the experimentally measured range of 
$\theta_{tilt}$, in agreement with experiment (blue square).
There is reasonable agreement between the predicted coincidence angle of 
$\theta_{tilt} = -4.5^{\circ}$
and that observed in the experiment of 
$\theta_{tilt} \approx -6.5^{\circ}\pm 0.5^{\circ}$, although we will shortly 
discuss the origins of this $2^{\circ}$ discrepancy.

To highlight the non-abelian dynamics
we have attempted to fit the observed data using the abelian Berry formula 
Eq. (\ref{preB}) instead of Eq.(\ref{prefactor}). Using Eqs. (\ref{preB}), 
(\ref{gg}) we repeat the same procedure described above and fit $\Phi_{AB}$ to
 the observed four coincidence angles for the orientations ($B_x > 0, B_y=0$) 
and ($B_y\ne 0, B_x=0$), as shown in Fig. 3c. The fitting parameters obtained 
are
 $g = 9.5$, $|\lambda| = 0.54, g_{xz} = +0.81$.
 These parameters were 
then used to predict the coincidence angles occurring for the orientation 
$B_x < 0, B_y=0$, shown in Fig. 3d. The key point is that the abelian theory 
always predicts three coincidence angles in contrast to the single coincidence 
observed in experiment.

Although the number of coincidence angles is not a topological invariant 
(for example it depends on the range of tilt angles available in the 
experiment), it is robust within both the non-abelian and the abelian 
theories. As shown in 
Appendix \ref{apG}, although the precise tilt angles at which the coincidences 
occurred are sensitive to the fitting parameters, the \emph{number} of 
coincidences could not be changed even after significant variation of the 
parameters.

Additional confirmation of the non-abelian dynamics comes from the
Dresselhaus interaction, neglected so far because of its
smallness.
In the co-rotating frame the spin-orbit coupling given by Eq.(\ref{b1})
 results in an energy splitting $\Delta_{\uparrow\downarrow}
=\omega_c\frac{\Phi}{\pi}$ between the ``up'' and ``down'' spin states,
see Eq.(\ref{prefactor}). The quantization axis for $\Delta_{\uparrow\downarrow}$
is tilted with respect to $z$.
The Dresselhaus interaction in the co-rotating frame takes the form
of a small periodic perturbation $ \sigma_z\cos\omega_ct$.
Since the quantization axis is tilted, this perturbation
drives transitions between the spin ``up'' and spin ``down''.
Because of the smallness of the perturbation the transitions are
significant only close to resonance,
$\Delta_{\uparrow\downarrow}\approx \omega_c$.
We use the amplitude of the Dresselhaus interaction as an additional fitting 
parameter, and find that it is close to the value known from the literature,
see Appendix \ref{apE}.
 The effect of the Dresselhaus perturbation is shown by the dashed curves in 
Figs. 3a,b. 
The tiny difference from the red and green solid curves in Fig. 3a, which 
do not include the Dresselhaus interaction, show that the effect of the 
interaction is very weak.
 On the other hand, for $B_x<0$
(Fig. 3b) the resonance condition
$\Delta_{\uparrow\downarrow}\approx \omega_c$ is satisfied
and the Dresselhaus term now becomes significant. This causes a clear 
difference between the solid and dashed blue curves in Fig. 3b, which 
completely removes the small disagreement between
experimental and theoretical values of the coincidence angle.

Of course, the inclusion of the Dresselhaus interaction does not influence the 
number of coincidence angles, which is a very robust number. Moreover the 
inclusion of the Dresselhaus term explains why the single coincidence 
for $B_x<0$ is 
not sharp, but occurs over a much wider range of angles than for 
$+B_x$ or $\pm B_y$ (seen as the slow phase inversion and small amplitude of 
the SdH oscillations in the range  $5^{\circ}<|\theta_{tilt}| < 10^{\circ}$ for 
blue traces in Fig. 2b).
This non-sharp transition for $B_x<0$ is explained by the inflection in the 
blue dashed curve in Fig.3b, which is due to the Dresselhaus interaction.
The ``inflection'' effect provides further confirmation of the non-abelian 
dynamics, since the small Dresselhaus perturbation is always
insignificant in the abelian theory.

Finally we present in Figs. 2b and c theoretical SdH curves calculated with 
modified
Lifshitz-Kosevich formula (\ref{osc}).
The agreement between theory (including Dresselhaus interaction) and
experiment is very good.
Overall, our data on the number of coincidences, supported by the
slow phase flip of the SdH oscillations for $B_x<0$,
provide unambiguous evidence for the non-abelian gauge field.

The non-abelian gauge field features centrally in theoretical proposals to 
exploit hole systems for spintronics and quantum information purposes, 
including the realization of the dissipationless spin Hall 
effect \cite{Murakami} and non-abelian manipulation of hole 
qubits \cite{Budich}. The capacity of hole systems in this context is 
further enhanced by the suppression of decoherence due to absence of the 
hyperfine interaction \cite{Chekhovich, Keane}. The observation of the 
non-abelian gauge field in a 2D hole system has positive implications for 
future studies of hole systems which rely on this concept.

\textbf{Acknowledgements}
We acknowledge Baruch Horowitz, Ulrich Zuelicke, Roland Winkler,
and Dimitry Miserev
for important discussions.


\appendix

\section{Derivation of the spin-orbit interaction for heavy holes}
\label{apD}

\begin{figure}[ht!]
 \includegraphics[width = 0.4\textwidth]{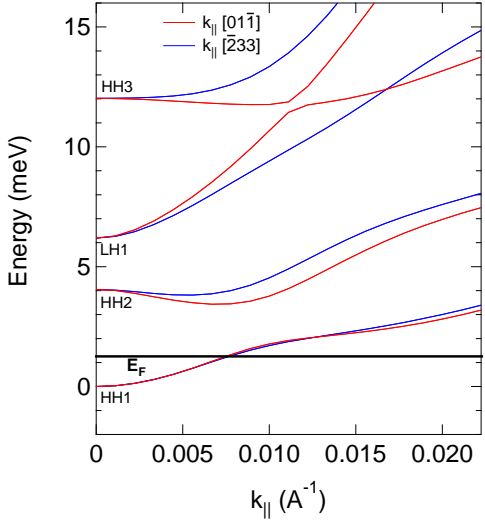}
\caption{Energy levels (2D dispersions) of holes in a (311) 
square quantum well of width $d=20$nm.
The Fermi level shown by the red horizintal line, $E_F\approx 1.3$meV,
corresponds to the hole density $n= 10^{11}\text{cm}^{-2}$.
}
\label{HH12}
\end{figure}

In a zinc blende semiconductor the hole wave function originates from
atomic $p_{3/2}$ orbitals resulting in an angular momentum $J=3/2$. In the 
long wavelength approximation, the effective Luttinger Hamiltonian for holes is
quadratic in the hole momentum ${\bm k}$~\cite{Luttinger}
(see also Ref. \cite{Winkler} 
\begin{eqnarray}
&&H_L = \left(\gamma_1 + \frac{5}{2} \gamma_2 \right) \frac{{\bf k}^2}{2 m_e} 
- \frac{\gamma_2}{m_e} \left(k_1^2 S_1^2 + k_2^2 S_2^2 + k_3^2 S_3^2  \right)
\nonumber \\
&&- \frac{\gamma_3}{m_e} \left(k_1 k_2 \{ S_1, S_2 \} + k_2 k_3 \{ S_2, S_3 \} 
+ k_3 k_1 \{ S_3, S_1 \} \right)\ ,\nonumber\\ 
 \label{Lut}
\end{eqnarray}
1, 2, 3 are the crystal axes of the cubic lattice,
$m_e$ is the electron mass,
$\{... \}$ denotes the anticommutator, and
$\gamma_1$, $\gamma_2$ and $\gamma_3$ 
are Luttinger parameters.
In GaAs 
$\gamma_1\approx 6.85$, $\gamma_2 \approx 2.1$, 
$\gamma_3 \approx 2.9$ \cite{gval}.
The Hamiltonian (\ref{Lut}) can be rewritten as
\begin{eqnarray}
H_L = \left(\gamma_1 + \frac{5}{2} \overline{\gamma}_2 \right)
\frac{{\bf k}^2}{2 m} 
- \frac{\overline{\gamma}_2}{m} \left({\bf k} \cdot {\bf S} \right)^2
+k_ik_jS_mS_nT^{(4)}_{ijmn}\ ,\nonumber
 \label{Lut1}
\end{eqnarray}
where
\begin{eqnarray}
\overline{\gamma}_2 = \frac{2 \gamma_2 + 3 \gamma_3}{5}\approx 2.6 \ .\nonumber
\end{eqnarray}
The irreducible 4$^{th}$ rank tensor $T^{(4)}_{ijmn}$ depends on the orientation 
of the cubic lattice, the tensor is proportional to $\gamma_3-\gamma_2$.
Neglecting $\gamma_3-\gamma_2$ compared to $\gamma_2$,
 the Luttinger Hamiltonian can be approximated by  the following
rotationally invariant (independent of the lattice orientation) Hamiltonian
\begin{equation}
\label{HL}
H_L\to H = \frac{\hbar^2}{2m_e} \left[ (\gamma_1 + \frac{5}{2} 
\overline{\gamma}_2) k^2 - 2\overline{\gamma}_2 (\boldsymbol{k} \cdot \boldsymbol{J})^2 \right] \ .
\end{equation}
Due to the 
confining potential $V(z)$, motion perpendicular to the 2D plane of the 
heterostructure is quantized, leading to the formation of 2D subbands, where 
only the lowest subband occupied in the low-temperature experimental regime. 
Assuming a square well confining potential of width $d$ we have 
$\langle k_z^2 \rangle = \frac{ \pi^2}{d^2}$. Since $\langle k_z^2 \rangle \gg k_F^2$, 
we may expand $ -(\boldsymbol{k} \cdot \boldsymbol{J})^2 = - k_z^2 J_z^2+ \dots$,
 with the leading term becoming diagonal in a basis of states with $J_z$. Due 
to the sign of the interaction, states with $J_z = \pm \frac{3}{2}$ 
(heavy hole) are lower in energy, and the splitting between these and states 
with $J_z = \pm \frac{1}{2}$ (light hole) at $k_x = k_y = 0$ becomes
\begin{equation}
\label{Ssplit}
\Delta_{hl} = 2\overline{\gamma}_2\frac{\pi^2  \hbar^2}{m_e d^2} 
\approx 9.6 \text{meV} \ \ .
\end{equation}
Here we take $d=20$nm.
The splitting between the lowest and the next heavy hole band 
at $k_x = k_y = 0$ is
\begin{equation}
\label{Hsplit}
\Delta_{h12} = \frac{3}{2}(\gamma_1-2\overline{\gamma}_2)
\frac{\pi^2  \hbar^2}{m_e d^2} 
\approx 4.6 \text{meV} \ \ .
\end{equation}
Numerical diagonalization of the full Luttinger Hamiltinian (\ref{Lut})
using the NextNano++ package~\cite{Birner}
gives the energy levels (2D dispersions) plotted in Fig. \ref{HH12}.
The HH1-HH2 splitting at $k_x = k_y = 0$ is pretty close
to (\ref{Hsplit}) while the HH1-LH1 splitting in Fig. \ref{HH12}
is somewhat smaller than (\ref{Ssplit}) because of the
$\sim (\gamma_3-\gamma_2)$ tensor corrections.
At hole density corresponding to our experiment,
$n\approx 10^{11}\text{cm}^{-2}$, only the lowest band is populated.
The lowest band dispersion $\epsilon({\bm k})$ enters Eq. (\ref{Heff}).
We describe this band by the effective spin $s=1/2$,
$|J_z=+3/2\rangle \equiv|\uparrow\rangle$,
$|J_z=-3/2\rangle \equiv |\downarrow\rangle$.
The Fermi momentum is  $k_F\approx 0.0079\AA^{-1}$ and
the Fermi energy, $E_F \approx 1.3$meV, is shown in Fig. \ref{HH12}
by the red horizontal line.
The heavy hole effective mass, $m=k\left(\frac{d\epsilon}{dk}\right)^{-1}$,
follows from Fig. \ref{HH12}.
At $k\to 0$ the mass is about $0.14m_e$
and at $k=k_F$ the mass is $m\approx 0.25m_e$. Obviously only the latter mass
is relevant to our analysis.

The off-diagonal part of $({\bm k}\cdot{\bm J})^2$ in the Hamiltonian 
(\ref{HL}),
$({\bm k}\cdot{\bm J})^2 \to \frac{1}{4}\left(k_-J_++k_+J_-\right)^2$, leads 
to heavy-light hole mixing.
\begin{eqnarray}
\label{Se}
&&|{\bm k},\uparrow\rangle=\left[|+\frac{3}{2}\rangle
+a k_+^2|-\frac{1}{2}\rangle \right]e^{i{\bm k}\cdot{\bm r}}\nonumber\\
&&|{\bm k},\downarrow\rangle=\left[|-\frac{3}{2}\rangle
+a k_-^2|+\frac{1}{2}\rangle \right]e^{i{\bm k}\cdot{\bm r}}\nonumber\\
&&a=\frac{\sqrt{3}\gamma_2 }{2m_e\Delta_{hl}}=
\frac{\sqrt{3}}{4\langle k_z^2\rangle}\ .
\end{eqnarray}
Taking the square well width $d=20$nm and the hole density 
$n= 10^{11}\text{cm}^{-2}$, we arrive at the following estimate for the 
mixing probability, 
$a^2k_F^4 =\frac{3}{4\pi^2}d^4n^2 \approx 1.2 \times 10^{-2}$. 
This very small mixing, of order $1\%$ in probability, is responsible for 
the SOC considered here.

The Zeeman interaction of a $J=3/2$ hole with magnetic field 
$\boldsymbol{B}$ is \cite{Winkler},
\begin{equation}
\label{dh}
\delta H=-\frac{g_0}{3}\mu_B
\boldsymbol{B} \cdot \boldsymbol{J}\ ,
\end{equation}
where $g_0\approx 7.2$.
Taking the matrix element of $\delta H$ between states Eq.(\ref{Se}) we find 
the effective matrix of $H_{soc}$
\begin{equation}
\label{Smatrix}
\langle \downarrow | H_{soc} | \uparrow \rangle  \equiv
\langle \downarrow | \delta H | \uparrow \rangle
 = -\frac{g_0 \mu_B}{4\langle k_z^2 \rangle}B_+ k_+^2  \ \ .
\end{equation}
Comparing this with $H_{soc}$ in Eq.(\ref{Heff}) we determine the coefficient 
$\alpha$ in this equation to be
\begin{equation}
\label{al}
\alpha=\frac{g_0 \mu_B}{4\langle k_z^2 \rangle}  \ \ .
\end{equation}
According to our fit of SdH data
 $|\lambda|=2|\alpha| k_F^2/\mu_B\approx 0.88$. Hence we find that 
$k_F^2/k_z^2 \approx 0.25$ and the probability of the heavy-light hole mixing 
is $a^2k_F^4 =\frac{3}{16}\frac{k_F^4}{\langle k_z^2\rangle^2} \approx 1.1 \times 10^{-2}$,
which is remarkably consistent with the estimate presented after Eq.(\ref{Se}).
It is worth noting that Eq.(\ref{al}) is approximate,
 since one should expect a comparable contribution to $\alpha$ which is not 
accounted for by the calculations presented.
So far, we have neglected the coupling to the vector potential created by 
$\boldsymbol{B}_{\parallel}$,
$(\boldsymbol{k} \cdot \boldsymbol{J})^2\rightarrow ((\boldsymbol{k} - e \boldsymbol{A}) \cdot \boldsymbol{J})^2$.
This coupling also gives a contribution to the coefficient $\alpha$, see 
Refs. \cite{Winkler, Uli}.
This contribution is highly sensitive to the exact shape of the confining 
potential and therefore cannot be reliably calculated \cite{Uli}. The 
kinematic form of $H_{soc}$ however remains unambiguous and we can fit the 
value of $\alpha$ to the experimental data.

\section{Sample and transport measurements}
\label{apA}

The $2$D hole system resides within a symmetrically doped $20$nm-wide
 GaAs/Al$_{0.33}$Ga$_{0.67}$As quantum well, grown on the low symmetry plane 
$(311)$ by molecular beam epitaxy. A heavily doped $n^{+}$ GaAs layer located 
$2.6\mu$m below the quantum well, acts as an \emph{in situ} back gate, 
allowing the 2D density to be tuned~\cite{Simmons}. At zero back-gate voltage, 
the density of the 2D hole system is $n=1.33\times 10^{11}cm^{2}$ with a 
corresponding mobility of $\mu=678,000cm^{2}V^{-1}s^{-1}$. Transport measurements
 were performed in a Kelvinox $100$ dilution refrigerator within the bore of 
a $15$T magnet at a base temperature of $25$mK, using standard lock-in 
techniques, with a constant ac current of $10$nA at a frequency of $5$Hz.
To perform tilted field measurements, the sample was mounted on a 
piezo-electric rotator which allowed for \emph{in situ} rotation to be 
conducted with an accuracy of $\pm 0.01^\circ$ \cite{Yeoh}.

\begin{figure}[ht!]
\includegraphics[width = 0.45\textwidth]{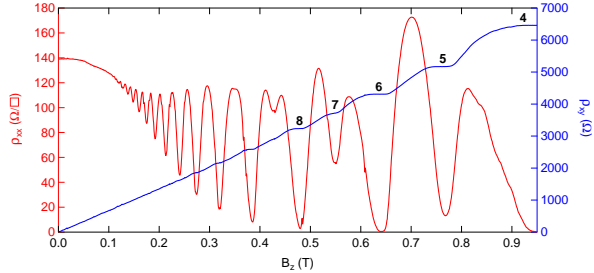}
\caption{
Plot of SdH oscillations $\rho_{xx}$ (in red) and corresponding Hall plateaus 
(in blue) as a function of perpendicular field $B_{z}$, taken at the symmetric 
operating point of $V_{BG}=+1.50\text{V}$, where the 2D carrier density is 
$n=9.26\times 10^{10}\text{cm}^{-2}$ and the mobility 
$600,000\text{cm}^{2}\text{V}^{-1}\text{s}^{-1}$.
}
\label{M1}
\end{figure}

Initially the 2D device was rotated to ${\theta}_{tilt}={90}^{\circ}$, so the 
magnetic field lies perpendicular to the sample plane, 
$B_{z}\ne 0$, $B_{\parallel} =0$, and the sample orientation confirmed by 
measuring the Hall plateaus as a function of perpendicular field, shown 
in Fig.\ref{M1} (blue).
The corresponding low-field oscillating longitudinal resistivity is shown in 
(red), with spin-splitting appearing for $B_{z}>0.35\text{T}$. For the purposes
 of our analysis we are only interested in low field data between 
$B_{z}=0.15$T and $0.25$T.

\section{Tilted field measurements}
\label{apC}

The coincidence method using tilted fields was first pioneered by Fang and 
Stiles in 1968 \cite{FangStiles} to study the Land\'{e} g-factor in 2D 
electron systems. Here we perform a similar set of tilted field transport 
measurements for a 2D hole system, taken along two crystal directions: 
the high symmetry $[0\bar{1}1]$ and the low symmetry $[\bar{2}33]$, as 
depicted in Fig.\ref{fig2}a. 
To achieve this, the device was first mounted on the rotator such that it 
tilts between the crystal axes $[311]$ and $[01\bar{1}]$, where the 2D plane 
is fully perpendicular to the field at $\theta_{tilt}=90^{\circ}$. The sample 
was then rotated towards the $[0\bar{1}1]$ direction till 
$\theta_{tilt}=+10^{\circ}$ to
introduce a parallel field component $B_{y}$, and the total field 
$\boldsymbol{B}$ swept, changing the sign of the in-plane field $\pm B_{y}$. 
This procedure was repeated for a number of different $+\theta_{tilt}$ with 
increasing in-plane field components. The experiment was then repeated for 
equivalent $-\theta_{tilt}$ and the results plotted in Fig.\ref{fig2}b. During 
a second cooldown, the sample was re-oriented to perform tilted measurements 
along the $[311]$ and low symmetry $[\bar{2}33]$ crystal axes. The experiment 
was then repeated for both $\pm \theta_{tilt}$ and the results shown in 
Fig.\ref{fig2}c.

\section{Tuning the confining potential with the back-gate voltage 
to compensate Rashba spin-orbit interaction}
\label{apB}

\begin{figure}
\includegraphics[width = 0.45\textwidth]{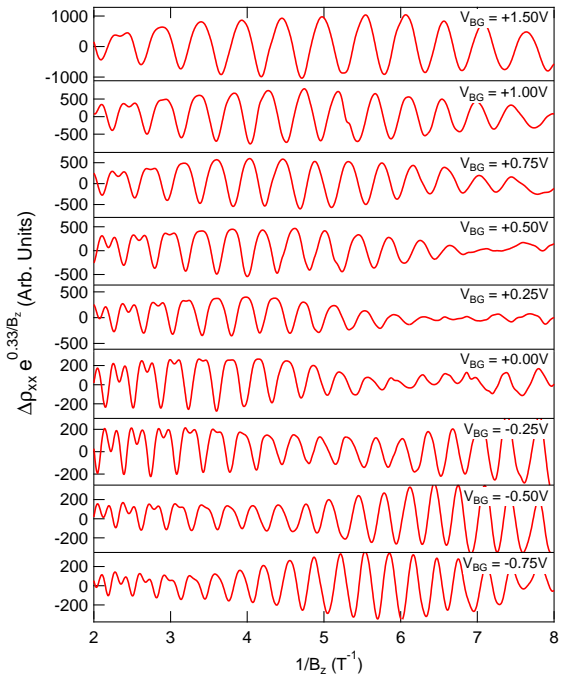}
\caption{
Plots of the SdH oscillations $\rho_{xx}$ periodic in inverse $B_z$, for 
different back-gate biases, with their amplitudes normalized by 
$e^{\frac{0.33T}{B_z}}$. The tilt angle $\theta_{tilt}=90^{\circ}$, so 
$B_{\parallel}=0$T. In the top panel at $V_{BG}=+1.50\text{V}$, the 2D carrier 
density is $n=9.26\times 10^{10}\text{cm}^{2}$ and increases to 
$n=1.53\times 10^{11}\text{cm}^{2}$ at $V_{BG}=-0.75\text{V}$ in the bottom 
panel. The back gate voltage $V_{BG}=+1.50\text{V}$, produces SdH oscillations 
with the least beating and hence this was selected as the operating point for 
the rest of the experiment.
}
\label{M2}
\end{figure}

The electric potential across the quantum well was tuned via the 
\emph{in situ} back gate, to adjust the confining potential. The presence of 
the Rashba SOC
results in beatings of the SdH oscillations even without any tilting of the 
magnetic field \cite{Eisenstein}. The Rashba interaction is sensitive to the 
back-gate voltage ($V_{BG}$), so by varying the applied bias-voltage, we can 
tune the system to minimize the amount of beatings and hence to eliminate the
Rashba interaction.
Fig.\ref{M2} shows these beatings in detail, where the SdH oscillations at 
each back gate voltage are periodic in $\frac{1}{B_z}$ and the amplitudes of 
these oscillations normalized for clarity by multiplying the datasets by 
$e^{\frac{0.33\text{T}}{B_z}}$ to remove the envelope. 
The data are taken without any tilting, $\theta_{tilt}=90^{\circ}$. From
Fig.\ref{M2} we select $V_{BG}=+1.50\text{V}$ as the final operating point 
with the least amount of beating in the SdH oscillations.
We will show that the major part of the Dresselhaus interaction does not
influence dynamics at $\theta_{tilt}=90^{\circ}$.  Hence, minimizing the beating
we tune the Rashba interaction to be close to zero.
This back-gate voltage is used as the operating point for the rest of the 
experiment. At this point the carrier density is 
$n=9.26\times 10^{10}\text{cm}^{-2}$ and the mobility is 
$600,000\text{cm}^{2}\text{V}^{-1}\text{s}^{-1}$.

\section{Sensitivity to Fitting Parameters}
\label{apG}

\begin{figure} [h!]
\vspace{10pt}
\centering
 \includegraphics[width = 0.45\textwidth]{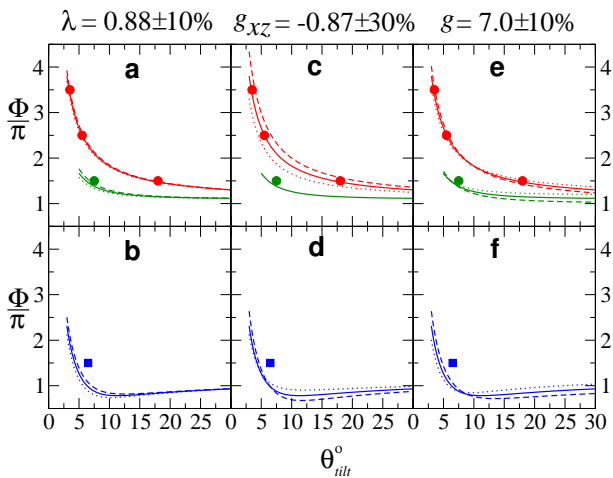}
\caption{
Comparison of the non-abelian theory 
(without accounting for the Dresselhaus interaction)
with experiment. The plot $\Phi/\pi$ [defined by the envelope of the 
resistivity oscillations $\rho_{xx} \propto \cos \Phi$, see Eq.(7)] as a 
function of tilt angle for varied parameters 
$\lambda, g_{xz}, g$.
 The top 
panels show the phase and the experimental coincidence points for 
orientations of the external field $B_x >0, B_y=0$ (red) and 
$B_y \ne 0, B_x=0$ (green). The bottom panels show the phase and the 
experimental coincidence points for the field orientation $B_x < 0, B_y=0$ 
(blue). Experimentally observed coincidence angles are shown in symbols. 
Solid lines corresponds to 
$\lambda = 0.88, g_{xz} = -0.87, g = 7.0$,
 and are 
identical to solid lines in Fig.3a,b. Dashed and dotted lines correspond to 
variations in $\lambda$ by $10\%$ (panels a, b), 
$g_{xz}$ by $\pm 30\%$ (panels c, d) and
$g$ 
by $\pm 10\%$ (panels e, f). Note 
there is only one green line in panel c since at $B_y \ne 0, B_x=0$ the phase 
is independent of $g_{xz}$.
}
\label{S1}
\end{figure}

The comparison of the experimental result with possible theories is 
presented in Fig.\ref{fig3}. Panels a and b show the non-abelian theory and 
panels c and d show the abelian theory. The non-abelian theory is consistent 
with experiment while the abelian theory is not consistent. Since the 
conclusions are based on our fits, a natural question which arises is 
`how sensitive 
is the number of coincidences with respect to variation in our fitting 
parameters?' 
In Fig.\ref{S1} we show the response of the non-abelian prediction, Eq.(7), 
as the fitting parameters are varied.
The layout and colour scheme are similar to Fig.3: the top panels show the
theoretical phase and the experimental coincidence points for orientations of 
the external field $B_x >0, B_y=0$ (red) and $B_y \ne 0, B_x=0$ (green). The 
bottom panels show the phase and the experimental coincidence points for the 
field orientation $B_x < 0, B_y=0$ (blue). Solid lines in Fig.\ref{S1} are 
identical to that in Fig.3a,b. In Fig.\ref{S1} panels a \& b correspond to 
$\pm 10\%$ variation of $\lambda$, the c \& d panels correspond to $\pm 30\%$ 
variation of $g_{xz}$, and the e \& f panels correspond to $\pm 10\%$ variation
of $g$.
 From these plots, the presented deviations are larger than those 
accepted in Fig.3. The curves corresponding to the lower boundaries of the 
parameters (dotted lines)
are too far away from the experimental points. On the other hand the curves 
corresponding to the upper boundaries of the parameters (dashed lines)
demonstrate an additional coincidence point (panels b,d,f)
which is not observed experimentally. This shows that the selected parameters, 
$|\lambda| = 0.88, g_{xz} = -0.87, g = 7.0$,
 provide the best fit to the data.
The curves in Fig.\ref{S1} do not account for the Dresselhaus 
interaction.
There is no point to account for the interaction for purposes of the
present analysis, since it hardly effects the red and green curves which are
used to determine the fit parameters, and it does not change the number of
coincidences.
Dresselhaus only deforms the blue curves in panels b, d, and f exactly 
in the same way as in the panel b of Fig. 3.\\

\begin{figure} [h!]
\vspace{10pt}
\centering
\includegraphics[width= 0.45\textwidth]{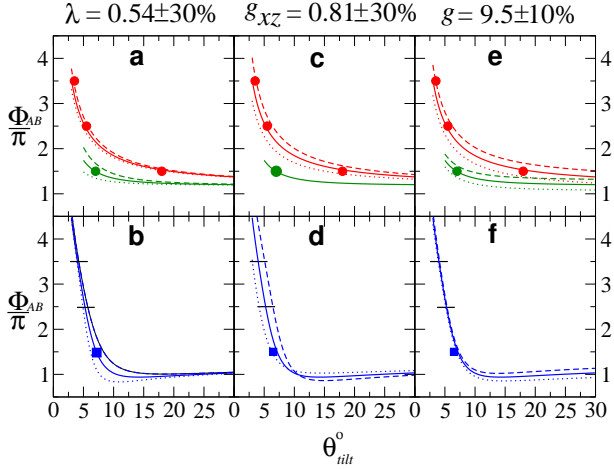}
\caption{
Comparison of the abelian theory with experiment. The plot $ \Phi_{AB}/\pi$ 
[defined by the envelope of the resistivity oscillations 
$\rho_{xx} \propto \cos \Phi$, see Eq. (8)] as a function of tilt angle for 
varied parameters 
$\lambda, g_{xz}, g$.
 The top panels show the phase and the 
experimental coincidence points for orientations of the external field $B_x >0,
 B_y=0$ (red) and $B_y \ne 0, B_x=0$ (green). The bottom panels show the phase 
and the experimental coincidence points for the field orientation  
$B_x < 0, B_y=0$ (blue). Experimentally observed coincidence angles are shown 
in symbols. Solid lines corresponds to 
$\lambda = 0.54, g_{xz} = 0.81, 9 = 9.5$,
 and are identical to the solid 
lines in Fig.3c,d.
 Dashed and dotted lines correspond to variation $\lambda$ by $30\%$ (panels 
a, b), $g_{xz}$ by $\pm 30\%$ (panels c, d) and $g$
by $\pm 10\%$ (panels e, f).
 The unobserved concidences are shown in panels b,d,f by short black horisontal
 lines. There is only one green line in panel c since at $B_y \ne 0, B_x=0$ 
the phase is independent of $g_{xz}$.
}
\label{S2}
\end{figure}

A similar comparison for the abelian theory, Eq.(8), is presented in 
Fig.\ref{S2}, where once again the parameters $\lambda, g_{xz}, m$ are varied. 
The top panels show the theoretical phase $\Phi_{AB}$ and the experimental 
coincidence angles for orientations of the external field $B_x>0, B_y=0$ (red) 
and $B_y \ne 0, B_x=0$ (green). The bottom panels show the phase and the 
experimental coincidence points for the field orientation $B_x<0, B_y=0$ 
(blue). Solid lines in Fig.\ref{S2} are identical to that in Fig.3c and 3d.
In Fig.\ref{S2} panels a and b correspond to $\pm 30\%$ variation of $\lambda$,
panels c and d correspond to $\pm 30\%$ variation of $g_{xz}$, and panels e and 
f correspond to $\pm 10\%$ variation of $g$.
 Despite the significant amount of
 variation in these parameters, the abelian theory always predicts three 
concidence angles for $B_x< 0, B_y=0$ (blue), whilst experimentally only one 
angle is observed. This discrepancy illustrates that the experimental data 
cannot be reconciled with the paradigm of Berry phases alone and renders our 
evidence for the non-abelian gauge field unambiguous.


\section{Accounting for the Dresselhaus interaction}
\label{apE}

Dresselhaus spin-orbit interaction arises due to the lack of inversion
symmetry in the bulk GaAs crystal.
The interaction is cubic in the  momentum $k$ and linear in the angular
momentum $J$, see Ref.~\cite{Winkler}.
In the coordinate system defined in Fig.\ref{fig2}a the leading
term of the Dresselhaus Hamiltonian is
\begin{equation}
\label{DH3}
H_D\to
-\frac{12\sqrt{22}}{121}b_D\langle k_z^2\rangle k_y\sigma_z =
-\frac{12\sqrt{22}}{121}b_D\langle k_z^2\rangle k_F\sigma_z\sin\theta \ .
\end{equation}
where $b_D=82$eV$\AA^3$, see Ref.\cite{Winkler}.
We neglect the subleading terms cubic in $k_{||}$.
The transformation (\ref{b1}) to the co-rotating frame  does not change
(\ref{DH3}).
Hence, in the co-rotating frame the Dresselhaus interaction works
as a weak periodic ``magnetic field'' superimposed on the constant
``magnetic field'' defined by Eq.(\ref{po}).
The projection of the periodic ``field'' on the direction
perpendicular to the direction of the constant ``field'',
\begin{equation}
\label{DH2}
 \sigma_z \sin\theta \to \sigma_{\perp} \ \frac{\alpha k_F^2B_{||}}
{\sqrt{ \left(\omega_c - \frac{ \Delta}{2}\right)^2
+ |\alpha k_F^2|^2 B_{||}^2 }} \sin\omega_c t \ ,
\end{equation}
generates spin flips.
Accounting for  the resonant part of the periodic perturbation one finds that
$\Phi$ given by Eq.(\ref{prefactor}) is replaced by $\Phi_D$,
\begin{eqnarray}
\label{prefactor1}
&&\frac{\Phi_D}{\pi}=1+\sqrt{\left(\frac{\Phi}{\pi}-1\right)^2+D^2}\\
&&D=D_0\frac{\alpha k_F^2B_{||}}
{\sqrt{ \left(\omega_c - \frac{ \Delta}{2}\right)^2
+ |\alpha k_F^2|^2 B_{||}^2 }}\ , \nonumber
\end{eqnarray}
where
\begin{eqnarray}
\label{D0}
D_0=\frac{12\sqrt{22}}{121} \frac{b_D \langle k_z^2\rangle k_F}{\omega_c} \ .
\end{eqnarray}
Without tilting, $\theta_{tilt}=90^{\circ}$, the Dresselhaus perturbation
(\ref{DH2}) is zero, $\Phi_D=\Phi$. Hence, our back-gate tuning performed
at $\theta_{tilt}=90^{\circ}$ does not compensate the Dresselhaus interaction.
The interaction becomes important at  intermediate values of  $\theta_{tilt}$.
The dashed curves in Figs. 3a,b display Eq.(\ref{prefactor1}) calculated
with $D_0=0.5$. We chose this value of $D_0$ to shift the coincidence angle
in Fig.3b from $\theta_{tilt} = -4.5^{\circ}$ to $\theta_{tilt} \approx -6.5^{\circ}$.
This is an additional fitting parameter.
Because of weakness of the Dresselhaus interaction the dashed and the solid
curves in Fig.3a are practically indistinguishable.
On the other hand, because of the resonance, the effect of the
Dresselhaus interaction in Fig.3b is significant.
We stress again the point made in the main text, the correct value
of $D_0$ necessarily leads to the extended range of $\theta_{tilt}$ over which the phase of the SdH oscillation inverts, clearly seen in the blue traces in Fig 2c.
The value of $D_0$ can be also calculated using Eq.(\ref{D0}). With
$B_z=0.2$Tesla and with parameters of the system discussed in the paper,
Eq.(\ref{D0}) gives $D_0\approx 0.8$. 
{
So, the Dresselhaus interaction is weak and insignificant
compared to the dominating  magnetic field controlled
spin orbit effects that drive the non-abelian
Berry phase.
The strength of the Dresselhaus interaction measured in our experiment
is slightly smaller but comparable with theoretical estimates presented
in Refs.\cite{Winkler,Durnev}.
}


\end{document}